\documentclass[pra,aps,twocolumn,superscriptaddress]{revtex4}
\usepackage{graphicx}
\usepackage{multirow}
\usepackage{amsmath}
\usepackage{amssymb}
\usepackage{dsfont}
\usepackage{amsthm}
\usepackage{eucal}
\usepackage{bm}
\usepackage{tikz}
\usepackage{upgreek}
\usepackage[final]{hyperref} % adds hyper links inside the generated pdf file
\hypersetup{
	colorlinks=true,       % false: boxed links; true: colored links
	linkcolor=blue,          % color of internal links
	citecolor=red,        % color of links to bibliography
	filecolor=magenta,      % color of file links
	urlcolor=black         
}
\usepackage{qcircuit}
\usepackage{pifont}
\usepackage{mathrsfs}
\usepackage{diagbox}
\usepackage{color}
\usepackage{mathtools}
\usepackage{latexsym}
\usepackage{float}

\theoremstyle{plain}

\theoremstyle{plain}

\newcommand{\ket} [1] {\vert #1 \rangle}
\newcommand{\bra} [1] {\langle #1 \vert}
\newcommand{\braket}[2]{\langle #1 | #2 \rangle}
\newcommand{\proj}[1]{\ket{#1}\bra{#1}}

\newcommand{\cmark}{\ding{51}}%
\newcommand{\xmark}{\ding{55}}%
\newcommand{\ts}{\textsuperscript}

\begin{document}
\title{ON states as resource units for universal quantum computation with photonic architectures}
\author{Krishna Kumar Sabapathy}
\email{krishna@xanadu.ai}
\affiliation{Xanadu, 372 Richmond St W, Toronto ON, M5V 2L7, Canada}

\author{Christian Weedbrook}
%\email{christian@xanadu.ai}
\affiliation{Xanadu, 372 Richmond St W, Toronto ON, M5V 2L7, Canada}

\begin{abstract}
Universal quantum computation using photonic systems requires gates whose Hamiltonians are of order greater than quadratic in the quadrature operators. We first review previous proposals to implement such gates, where specific non-Gaussian states are used as resources in conjunction with entangling gates such as the continuous-variable versions of C-PHASE and  C-NOT gates. We then propose  ON states which are superpositions of the vacuum and the $N^{th}$ Fock state, for use as non-Gaussian resource states. %The motivation is  to identify basic units of resource states requisite for universal quantum computation.
We show that ON states can be used to implement the cubic and higher-order quadrature phase gates to first order in gate strength. There are several advantages to this method such as reduced number of superpositions in the resource state preparation and greater control over the final gate.  We also introduce useful figures of merit to characterize gate performance. Utilising a supply of on-demand resource states one can potentially scale up implementation to greater accuracy, by repeated application of the basic circuit.   
\end{abstract}

%\pacs{03.67.Mn, 03.65.Yz, 42.50.Dv}
\maketitle

\section{Introduction}
The physical medium of photonics is considered a promising candidate for scalable and robust implementation of quantum information processing \cite{rmp,rmpbl,flbook}.  There have been various experimental advances and proposals to this end \cite{photonics1,photonics2}, especially in the field of integrated photonic circuits \cite{ipc1,ipc2,ipc3,ipc4}. Substantial research continues to focus on the role of photonics in universal quantum computation \cite{uni1,uni2,uni3,uni4,daiqin2018} and in demonstrable quantum computational advantage \cite{ashley,xanadu1,xanadu2}. Further, photonic systems are also a suitable physical medium for fault-tolerant quantum computation \cite{gkp,nicmenerror}.

It is well known that a basic requisite set of gates for universal continuous-variable quantum computation are ones with Hamiltonians given by $\hat{x}, \hat{x}^2, \hat{x}^3$, along with a two-mode Gaussian gate and a Fourier gate \cite{seth}. So the cubic phase gate plays a pivotal role since its the lowest order non-Gaussian gate in this elemental tool kit that provides an entry into universal quantum computation. One concrete application using the cubic and quartic gates is the simulation of the Bose-Hubbard model \cite{uni2}. Non-Gaussianity in general has been an active area of recent interest since it not only plays a role here in quantum computation but has also proven advantageous in other quantum information processing tasks such as parameter estimation \cite{paraest}, generation of entangled states \cite{ent1,ent3,ent4}, quantum communication \cite{pang} and teleportation \cite{tele1,tele2,tele4}.

There are already several proposals for implementation of the cubic phase gate each with its advantages and disadvantages. We have identified four broad approaches that provide approximate schemes to implement the cubic phase gate \cite{gkp,atf,rus,angm}. It is important to note that the methods focused primarily on the implementation of the cubic phase gate to first order in its Taylor expansion (or equivalently a weak cubic gate). 

Any general quadrature phase gate is of the form $\widehat{\Theta}(\gamma) = \exp{[i\gamma x^N]}$, where $|\gamma|$ is the gate strength and $N$ is the order of the gate. Since we are interested in implementing these gates approximately, we consider its Taylor expansion $\widehat{\Theta}(\gamma) = \sum_{m=0} (i\gamma \hat{x}^N)^m/m!$. By accuracy, we denote the power of the gate strength in the expansion up to which the gate is being approximated. For example, first order in accuracy of $\widehat{\Theta}(\gamma)$ is the expansion $1 + i\gamma x^N$, and so on. It is thus the triad of gate strength $|\gamma|$, order of the Hamiltonian $N$, and the accuracy of the Taylor expansion, that plays an important role in the description and implementation of these quadrature phase gates.  Note that for very small gate strengths, low accuracy would approximate the gate well.

For Gaussian elements ($N=1,2$) one can implement the gates to very high accuracy for all gate strengths. However, for the cubic gate, implementation to even the first order in accuracy has been a challenge. To generate quartic (and higher-order) phase gates, one needs to use various gate approximations and concatetation methods \cite{suzuki1,seth,sefi1,sefi2}, and this is where the accuracy plays an important role.

%\begin{figure}
%\begin{center}
%\scalebox{0.75}{~~~~~~\includegraphics[width=\columnwidth]{orderlat2.eps}}
%\end{center}
%\caption{Order-accuracy lattice. In this figure we present a visualization of current state-of-the-art for implementing continuous-variable quadrature phase gates which are of the form $\Theta(\gamma)= \exp(i\gamma\hat{x}^N)$. The x-axis (Accuracy)  denotes the order of the Taylor series expansion of $\Theta(\gamma)$ in parameter $\gamma$ and the y-axis (Order) denotes the order $N$ of the Hamiltonian. For $N=1$ and $N=2$ which corresponds to Gaussian elements, the corresponding gates are realizable to very high accuracy experimentally (green) for all gate strengths. However, this takes a dramatic turn for $N=3$ where even realizing the operator to first order expansion in $\gamma$ has been a challenge (orange). We explore the cubic phase gate to higher accuracy and the quartic gate to first order in $\gamma$. All higher-order polynomials and more accurate implementations are some of the future directions (red). The reason this is important is because while for Gaussian gates one has access to the unitaries directly, for non-Gaussian gates the current methods indicate that one has to construct the gate order-by-order.}
%\label{fig0}
%\end{figure}

Keeping the order-strength-accuracy triad in mind we emphasize that the cubic phase gate is sufficient for universal quantum computation along with a (non-unique) minimal set of Gaussian elements only when the cubic gate is implementable to sufficiently high accuracy for all gate strengths, assuming repeated applicability. In this article we explore the optical implementation of the cubic phase gate to higher accuracies and also the quartic gate to first order in accuracy. 

The outline of the paper is as follows. In Section II we provide a brief overview of the cubic phase gate and its closely related companion, the cubic phase state . In Section III we review four broad schemes that we have identified under which previous implementations can be classified. In Section IV we introduce our implementation of the cubic and quartic gates using ON states, along with its basic properties. We conclude in Section V.  

\section{Cubic phase state and cubic phase gate}
The cubic phase gate plays a crucial role for universal quantum computation since it has the lowest order Hamiltonian among non-Gaussian quadrature phase gates. Hence the cubic phase gate $V(\gamma)$ and the  related cubic phase state $\ket{\gamma}$  have received substantial attention. These are defined as  
\begin{align}
V(\gamma) &= e^{i \gamma \hat{x}^3}, \\
\ket{\gamma} &=  V(\gamma)\ket{0}_{\rm p} = \delta \int dx\, e^{i \gamma x^3} \ket{x}, \label{custate}
\end{align}
where the subscript p denotes a momentum eigenket and $\gamma \in \mathbb{R}$.  Note further that the scalar $\delta$ is only ornamental since the state is  not normalizable. We also wish to point out that we only deal with a single mode of an electromagnetic field in the entire article, and we work in natural units where $x$ is dimensionless and we further set $\hbar =1$. 

\begin{figure}[ht!]
\begin{center}
\scalebox{0.47}{\includegraphics{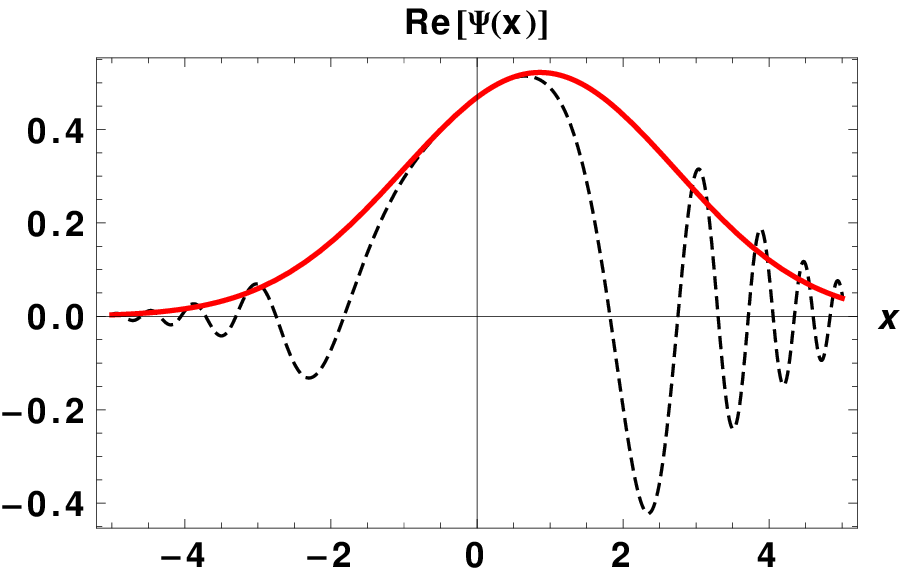}\hspace{0.2cm}\includegraphics{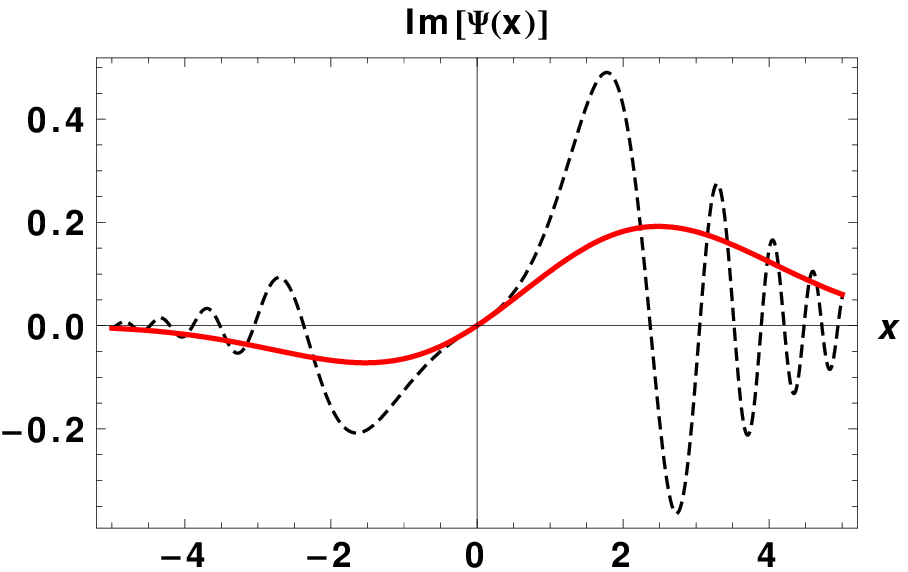}}
\end{center}
\caption{Real (left) and imaginary (right) parts of the  wave-function (thick red line)  of a squeezed displaced vacuum state $\ket{\psi}$  and its modulation (dashed line) by the action of the cubic phase gate. }
\label{figa}
\end{figure}

We briefly recall a few basic properties that we use later. The action of the cubic gate at the level of the position wave-function is 
\begin{align}
V(\gamma)\ket{\psi} = \int dx\, \psi(x) e^{i\gamma x^3} \ket{x} \nonumber\\
\Rightarrow \psi(x) \to \psi(x) e^{i\gamma x^3}.
\end{align}
So we see that the wave-function is modulated by a position dependent phase, so the oscillations become very rapid for large $x$ as shown in Fig. \ref{figa}. However, the probability amplitudes $|\psi(x)|^2$ remain unchanged under the action of the cubic gate.  
In some realistic settings where resource states are used in a basic gate-teleportation type circuit the ket $\ket{0}_p$ is replaced by a suitable squeezed state. The Wigner function of the ideal cubic state is given by \cite{ghose-sanders}
\begin{align}
W(x,p) = N_0\, {\rm Ai}\left(b_0 [3\gamma x^2 -p] \right),
\end{align}
with parabolae $3 \gamma x^2 - p = {\rm constant}$ being contours in phase space. Here, $N_0 = 2\pi \delta |b_0|$, $b_0 = (4/3\gamma )^{1/3} $, and ${\rm Ai}$ stands for the Airy function.  The cubic phase gate in the Heisenberg picture induces the following transformation on the quadrature operators 
\begin{align}
V(\gamma)^{\dagger}\, \hat{x} \,V(\gamma) &= \hat{x}, \nonumber\\
V(\gamma)^{\dagger}\, \hat{p} \,V(\gamma) &= \hat{p} +  3 \gamma \hat{x}^2.
\end{align}
Since we are interested in approximations to the cubic phase gate and state, it is useful to express them as Taylor expansions in the parameter $\gamma$ given by 
\begin{align}
V(\gamma) &= 1\!\!1 + i \gamma \hat{x}^3 -\gamma^2 \hat{x}^6/2 + O(\gamma^3), \nonumber\\
\ket{\gamma} &=  \int dx \, (1 + i\gamma x^3 + O(\gamma^2)) \ket{x} \nonumber\\ 
&= \ket{0}_p + i \gamma N^{'} \int dx \, x^3 \ket{x} + O(\gamma^2),
\end{align}
for some normalization $N^{'}$. Note that the approximate phase state has a cubic imaginary component superposed with an infinitely squeezed vacuum state when expanded to first order in $\gamma$. \\

\section{Review of previous techniques}
Previous routes to non-linear gates can be classified into four broad approaches which we recapitulate now in no particular order.  The first is the original approach by Gottesman-Kitaev-Preskill (GKP) \cite{gkp} where they reduced the problem of generating a cubic phase gate to that of a cubic phase state. GKP then provided an approximate scheme to generate this resource state using two-mode squeezed states, displacements, a photon-number resolving detector, and squeezing. It is useful to introduce notation for two-mode entangling gates where $C_X = e^{-i\hat{x}_1 \hat{p}_2}$ (controlled-X), $C_Z=  e^{i\hat{x}_1 \hat{x}_2}$ (controlled-Z), $C_{\alpha} = e^{i\hat{x}_1 (\alpha \hat{a}_2^{\dagger} -\alpha^* \hat{a}_2)}$ (controlled-$\alpha$),  and single-mode displacements $X(a)=e^{-ia \hat{p}}, \, Z(a)=e^{ia \hat{x}}$. The $C_{\alpha}$ gate is also sometimes referred to as the QND gate, but we choose the former for clarity. We spend more time in the description of the GKP circuit as this was the first method in this topic and it will also be useful for us.\\ %There are several metho(ads that try and improve or provide alternative approximates to this resource state such as the a squeezing applying to a state that is in a superposition of the 0-1-3 fock states (MFF) and the photon-subtracted squeezed displaced state (ATF1). \\

\noindent \textbf{GKP circuit with a resource state}\,\cite{gkp,ghose-sanders,gkpstates,gkperror}. Take an input state $\ket{\psi}$ in a tensor product with a resource state $\ket{\phi_r}$. Let $\ket{\phi_r} = \int dx\, \phi_r(x) \ket{x}$ with position wavefunction $\phi_r(x)$. Now apply the gate $C_X^{\dagger} = e^{i \hat{x}_1 \hat{p}_2}$. Finally, perform a position homodyne measurement $\Pi_x$ of quadrature $\hat{x}$ on the resource mode to obtain
\begin{align}
\label{eq9}
\Qcircuit @C=1.em @R=1em {
&&&&C_X^{\dagger}
&&\\
\lstick{\ket{\psi}} &\qw &\qw&\qw&\ctrl{3} & \qw &\qw&  \push{~~\widehat{T}_1(q) \ket{\psi}\rule{-3em}{1em}} \qw \\
&&&&&&\\
&&&&&&\\
\lstick{\ket{\phi_r}} & \qw&\qw&\qw&\targ & \qw&\measureD{\mbox{$\Pi_x$}} & \cw ~~~~q ,~
}
\end{align}
where a conditional operator $\widehat{T}_1(q)$ resulting from a measurement outcome $q$ gives an output state \begin{align}
\label{t1q}
\widehat{T}_{1}(q) \ket{\psi} &= \bra{q}  C_X^{\dagger} \ket{\psi} \ket{\phi_r} \nonumber\\
&= \bra{q}  \int dx dy \,  \psi(x)  \phi_r(y)  e^{i \hat{x}_1 \hat{p}_2} \ket{x} \ket{y} \nonumber\\
& =  \bra{q}  \int dx dy \, \psi(x) \phi_r(y) \ket{x} \ket{y-x} \nonumber\\
&=  \int dx \,  \psi(x) \phi_r(x+q) \ket{x} \nonumber\\ 
&=  \int dx \,  \psi(x) \phi_r(\hat{x}+q) \ket{x} \nonumber\\ 
&= \phi_r(\hat{x}+q) \ket{\psi}.
\end{align}

In effect the circuit in Eq. \eqref{eq9} implements the operator $\widehat{T}_1(q)$ with probability 
\begin{align}
p(q) &= \bra{\psi} \phi_r(\hat{x}+q)^{\dagger} \phi_r(\hat{x}+q) \ket{\psi}\nonumber\\
&= \int dx dy \bra{x} \psi(x)^* \phi_r(x+q)^* \phi_r(y+q) \psi(y) \ket{y}\nonumber\\
&= \int dx |\psi(x)|^2 |\phi_r(x+q)|^2.
\label{prob}
\end{align}
The operation $\widehat{T}_1(q)$ is a filter, i.e., non-unitary. If the resource state is non-Gaussian, then so is $\widehat{T}_1(q)$. The key idea is to interpret this filter operation as an approximate non-linear gate.\\

%\begin{figure}
%\begin{center}
%\includegraphics[width=\columnwidth]{gkp.eps}
%\end{center}
%\caption{basic elements}
%\label{fig1}
%\end{figure}

\noindent \textbf{GKP with ideal cubic state.} 
Let us now use as a resource an ideal cubic state of Eq. \eqref{custate} into the GKP circuit in Eq. \eqref{eq9}. Then we have that 
\begin{align}
\widehat{T}_1(q) \ket{\psi} = e^{i \gamma (\hat{x}+q)^3} \ket{\psi}.
\end{align}
One needs to correct for the shift in the cubic factor in the applied operator that resulted from the measurement. This can be achieved by a Gaussian feed-forward operator  $\widehat{F}_q = \exp{[-i\gamma(3 \hat{x}^2 q + 3 \hat{x}q^2 + q^3)]}$, so that 
\begin{align}
\widehat{T}_q = \widehat{F}_q \widehat{T}_1(q) \ket{\psi} = V(\gamma) \ket{\psi}, 
\label{gkpcorr}
\end{align}
which was the required aim. Here $\widehat{F}_q = P(-3\gamma q) Z(-3\gamma q^2)$, with $P(t) = e^{i t \hat{x}^2}$, $Z(t) = e^{i t \hat{x}}$, and the overall phase is immaterial. Therefore, these corrections take the form of the optical elements of squeezing and rotations along with displacements, respectively.\\
%\noindent \textit{Note.} If we average over the outcomes of the homodyne we obtain the state $\tilde{\rho} = \int_\Omega dq\, p(q) \ket{\chi_q} \bra{\chi_q}$, where $\Omega$ encodes some detection error region for the homodyne measurement. This may play a role in experimental considerations.\\

\noindent\textit{Kraus operators.} In the language of quantum channels, what the GKP circuit implements is one Kraus operator of a channel where the system-ancilla unitary is the $C_x^{\dagger}$ gate, the ancilla state is the resource state $\ket{\phi_r}$, and the homodyne measurement is the basis in which the ancilla is measured. Conditioned on a particular measurement outcome, the corresponding Kraus operator is applied to our input state. The output state is then sent through a correction optical setup that depends on the outcome of the homodyne measurement to obtain the action of the cubic phase gate. \\ 

\noindent \textit{GKP with realistic resource state.}  
GKP provided a realistic procedure for the resource state  using the following optical circuit
\begin{align}\label{eq:GKPstate}
 \mbox{
\Qcircuit @C=0.5em @R=2em { 
\lstick{\ket{0}}& \qw&\gate{S}& \multigate{1}{B(\pi/4)} & \qw & \gate{Z(w)}&\measureD{\mbox{$\Pi_{n}$}} & \ustick{m}  \controlo \cw  \cwx[1] \\ 
%&&\ghost{B(\pi/4)} &&&\\
\lstick{\ket{0}}& \qw&\gate{S^{-1}}&\ghost{B(\pi/4)} & \qw & \qw&\qw &\gate{S(m)} & \rstick{\ket{\phi_r} ~.} \qw  }}
\end{align}
The state just after the beam splitter action is the two-mode squeezed state, $Z(w)$ is a displacement,  $\Pi_n$ is the photon number detection, and $S(m)$ is a measurement dependent squeezing that has to be applied. In the limit of large squeezing, displacement, and photon number detection $z,w,m$, one recovers the approximate cubic phase state. However, an analysis in \cite{ghose-sanders} places these large limit requirements to be experimentally challenging.\hfill $\square$\\

Apart from the GKP prescription for their realistic resource states, there are other resource states that have been considered for use in the GKP circuit in literature. One such example (we denote Marek-Filip-Furusawa or MFF) used a 0-1-3 Fock superposition state as in Refs. \cite{mff} and \cite{emulating}. Another approach which considered a variation of the original GKP circuit (we denote Arzani-Treps-Ferrini1 or ATF1 \cite{atf})  considered a single photon-subtracted squeezed displaced state $\hat{a} S(r) D(\alpha) \ket{0}$, that is also related to a weak cat state \cite{cat,cat2}. In this method a measurement-dependant monomial of the form $(\hat{x}-\lambda)$ is implemented along with an extra Gaussian damping factor.  \\

\noindent \textbf{Arzani-Treps-Ferrini2 (ATF2).} The second approach in Ref. \cite{atf} involves a photon counter used in place of the homodyne measurement in a GKP-type circuit along with a squeezed displaced state as a resource, as shown in the circuit 
\begin{align}\label{atfeq}
\mbox{
\Qcircuit @C=1.em @R=1em {
&&&&C_Z\\
\lstick{\ket{\psi}} &\qw &\qw&\qw&\ctrl{3} & \qw &\qw&  \push{~~I_j(\hat{x}) \ket{\psi}\rule{-3em}{1em}} \qw \\
&&&&&&\\
&&&&&&\\
\lstick{\ket{\phi_r}} & \qw&\qw&\qw&\ctrl{0} & \qw&\measureD{\mbox{$\Pi_n$}} & \cw~~~~~j ~~.
}
}
\end{align}
The effective filter operator implemented by the circuit can be obtained from 
\begin{align}
I_j(x) = {\cal F}_{y\to x}[\phi_r(y) n_j(y)],
\end{align}
where ${\cal F}$ is the Fourier transform and $n_j(y)$ is the position representation of the $j$ photon excited vacuum state.  Then using a single photon detection, and applying a suitable displacement to the outcome, the circuit implements an exact monomial operator of the form $(\hat{x}-\lambda)$. This is along with a Gaussian damping operator which is a byproduct of finite squeezing that one cannot avoid. To implement the cubic phase gate to first order, the circuit needs to be implemented thrice with suitably chosen resource state parameters for each repetition.  \\

%\begin{figure}
%\begin{center}
%\includegraphics[width=\columnwidth]{atf2.eps}
%\end{center}
%\caption{resource state is a squeezed displaced state and the measurement outcome is single fock state. $I(x) = {\cal F}[\phi_r(x) \psi_j(x)]$ where $\psi_n(x)$ is the position representation of the $n$th fock state and ${\cal F}(\cdot)$ represents the Fourier transform.}
%\label{fig3}
%\end{figure}

\noindent \textbf{Marshall-Pooser-Siopsis-Weedbrook (MPSW).} This is also known as the `repeat-until-success' method \cite{rus}). This method also implements a monomial operator $U_{\ell} = 1 + \lambda_{\ell} \hat{x}$, which is of a different form compared to previous methods, using an intricate measurement in the circuit given by
\begin{align} \label{eqrus}
\mbox{
\Qcircuit @C=1.em @R=0.8em {
&&&&C_{\beta}
&&\\
\lstick{\ket{\psi}} &\qw &\qw&\qw&\ctrl{4} & \qw &\qw&  \push{~~U_{\ell}\ket{\psi}\rule{-2em}{1em}} \qw \\
&&&&&&\\
&&&&&&\hspace{2.5cm}\otimes\\
&&&&&&\\
\lstick{\ket{\alpha}} & \qw&\qw&\qw&\gate{D(\beta)} & \qw&\measureD{\mbox{QND}} & \cw ~~~~ \ket{1},
}
}
\end{align}
where $C_{\beta} = e^{i\hat{x}_1 (\beta \hat{a}^{\dag} - \beta^* \hat{a})}$. 
Then to realize the first order cubic gate, the circuit is implemented thrice with suitably chosen parameters $\lambda_{\ell}$ of the unitary. This method has some advantages such as a simple resource state and no dynamic feed-forward elements. The QND  denotes a   conditional measurement outcome corresponding to the projection operator $P_{\bar{0}} = 1\!\!1 - \proj{0}$.  \\

%\begin{figure}
%\begin{center}
%\includegraphics[width=\columnwidth]{rus.eps}
%\end{center}
%\caption{}
%\label{fig4}
%\end{figure}

\noindent \textbf{Adaptive non-Gaussian measurement (AnGM).} The final method \cite{angm} utilizes a non-Gaussian ancilla state embedded in an adaptive heterodyne measurement (dashed box) in the circuit given by 
\begin{align}\label{eqangm}
\mbox{
\Qcircuit @C=0.3em @R=3em { 
\lstick{\ket{\psi}}&   \qw &\multigate{1}{BS}&\qw&\qw&\qw&\qw&\qw& \qw &  \gate{X}& \rstick{\widehat{T} \ket{\psi}\, ,~~~~} \qw\\
\lstick{\ket{0}_x}&  \qw &\ghost{BS} &\qw &\qw&\multigate{1}{BS}&\qw&\measureD{\Pi_x}&\controlo \cw&\cwx&\\
\lstick{\ket{\phi_r}}&  \qw &\qw&\qw & \qw& \ghost{BS}&\qw&  \qw& \measureD{\Pi_{p(\theta)}}  \cwx&  \control \cw \cwx \gategroup{2}{6}{3}{9}{1.4em}{--}&\\
}}
\end{align}
where   $\ket{0}_x$ denotes a position ket and $\ket{\phi_r}$ is a resource state.
The result of the first homodyne measurement on $\hat{x}$ is fed into the rotation value $\theta$ of the second quadrature $\hat{p}_{\theta}$. Choosing a suitable resource state that is a superposition of the first four Fock basis states implements the cubic gate to first order. This method is attuned to one-way quantum computation based on cluster architectures as already proposed in Ref. \cite{bilayer}. \hfill $\square$

%\begin{align}\label{eqangm}
%\mbox{
%\Qcircuit @C=0.2em @R=2.5em { 
%\lstick{\ket{\psi}} &\qw\qw\qw&\multigate{1}{BS} \qw \qw &\qw&\qw&\qw&\qw&\qw&\qw&\qw &\qw&\qw&\gate{X}&\qw&~~~~~~~~\widehat{T}\ket{\psi}~~\\ 
%\lstick{\ket{0}_x}&\qw\qw\qw&\ghost{BS} \qw \qw &\qw&\qw&\qw&\qw &\qw&\multigate{1}{BS}& \qw&\qw&\measureD{\Pi_{p(\theta)}} &\control \cw \cwx \\
%&& &\lstick{\ket{\phi_r}}&\qw&\qw&\qw&\qw\qw &\ghost{BS} &\qw &  \measureD{\Pi_x}  \gategroup{2}{9}{3}{13}{1.4em}{--} &\control \cw \cwx  &&\\ 
%}}
%\end{align}
%\begin{figure}
%\begin{center}
%\includegraphics[width=\columnwidth]{ngm.eps}
%\end{center}
%\caption{}
%\label{fig4}
%\end{figure}

Each of the above mentioned methods have different advantages and disadvantages both from a theoretical and an experimental point of view. But all the methods roughly fall under some form of ancilla-driven quantum computation. We present a comparison of the different methods toward the end of the article in Table. \ref{table2} after presenting our own in the next section. %While the first three methods deal with the traditional circuit (gate) based quantum computation model, the last method is more suitable for measurement based quantum computing using cluster states). %Our approach in this article is to look at the underlying elements of the various schemes and simplify various steps in the experimental proposal to improve flexibility of the constituent elements to potentially increase the chances of feasibility of the cubic phase gate. 

\section{GKP circuit with ON states as resources}
In this Section we study the use of ON resource states for implementing higher-order quadrature phase gates.  We define ON states as superpositions of vacuum and the $N$\ts{th} Fock state i.e.,
 \begin{align} \label{onstate}
 \ket{ON} = \frac{1}{\sqrt{1+|a|^2}} (\ket{0} + a \ket{N}).
 \end{align}
%As an example we plot the Wigner function of the ON state in Eq. \eqref{onstate} with $N=3$ and $a=0.5$ in Fig. \ref{onfig}. 
%\begin{figure}[ht!]
%\begin{center}
%\scalebox{0.2}{\includegraphics{on-wigner.png}}
%\end{center}
%\caption{Wigner function of the ON state in Eq. %\eqref{onstate} with $N=3$ and $a=0.5$.}
%\label{onfig}
%\end{figure}
Our approach using ON states falls in the first method of using a resource state in a GKP circuit. In the case of the 03 state (i.e., an ON state with $N=3$) we implement directly the first order of the cubic operator with a Gaussian damping as opposed to the monomial of the methods of ATF1, ATF2 and MPSW. Also the 03 state has already been experimentally generated \cite{uptothree}. We only require a superposition of two Fock states as opposed to three or four Fock states as required by the other methods. We also present a  clear demonstration of the quartic gate to first order as proof-of-principle since the required 04 state is experimentally challenging, but potentially possible by extending the existing methods. 
Focusing on the cubic phase gate, we discuss the effective operator being implemented, the experimental preparation of the resource state, and provide clear gate performance figures of merit. 
 
\subsection{Cubic gate}
\noindent \textbf{GKP circuit with the 03 state.} We define the 03 state as 
\begin{align}
\ket{\phi_r} \equiv \ket{\psi_a} = c_a (\ket{0} + a\ket{3}),~ c_a= (1+|a|^2)^{-1/2}. 
\end{align}
The position wave function of this state is given by
\begin{align}
\psi_a(x) &=  c_a\pi^{-1/4} e^{-x^2/2} \left[ H_0(x) + \frac{a}{\sqrt{3! 2^3}} H_3(x)\right]  \nonumber \\ 
&= c_a\pi^{-1/4}  e^{-x^2/2} \left[1 + \frac{2a}{\sqrt{3}}(x^3- 3x/2) \right]. 
\end{align}
We set $a = i \sqrt{3}a_0/2$ where $a_0 \in \mathbb{R}$, then the above equation reduces to 
\begin{align}
\psi_a(x) &= c_a\pi^{-1/4} e^{-x^2/2} \left[ 1 + i a_0 (x^3 - 3x/2)\right]. 
\end{align}
Then the  resulting  operator from using the 03 state as a resource state in the GKP circuit  conditioned on the homodyne outcome $q$ is given by Eq. \eqref{t1q} (we drop the constants since the effective operator is anyway only a filter operation) to be
\begin{align}
\widehat{T}_1(q) = e^{-(\hat{x}+q)^2/2} \left[ 1 + i a_0 ((\hat{x}+q)^3 - 3(\hat{x}+q)/2)\right]. 
\label{t1q03}
\end{align}
Let us now assume that $a_0<<1$, then we can approximate the terms in the square bracket in Eq. \eqref{t1q03} as a unitary operator to obtain
\begin{align}
&\widehat{T}_1(q)= e^{-i3a_0 q/2} \hat{A}_q \, Z(-3 a_0/2)\, e^{i a_0 H}, \text{where }  \nonumber \\
&\hat{A}_q = e^{-(\hat{x}+q)^2/2}, H = (\hat{x}+q)^3, Z(s) = e^{is \hat{x}}.
\end{align}
Note that $\hat{A}_q$ is not a unitary but a damping non-trace-preserving Gaussian noise operator that must be accounted for. The action of $\hat{A}_q$ on any state $|\chi\rangle$ is given by 
\begin{align}
\hat{A}_q |\chi\rangle = \hat{A}_q \int dx |x \rangle \langle x | \chi\rangle =  \int dx e^{-(x+q)^2/2} \chi(x) |x\rangle.
\label{aqx}
\end{align}
Further, the output of the circuit needs to be normalized to obtain the corresponding state. We can correct the Hamiltonian $H$  by using a feed-forward setup which consists only of Gaussian elements as already mentioned in Eq. \eqref{gkpcorr}. Conditioned on the homodyne measurement $q$ we perform the following correction given by 
\begin{align}
\widehat{F}_G = e^{i3a_0 q/2} e^{-ia_0(3\hat{x}^2q + 3 \hat{x} q^2 + q^3)} \, Z(3a_0/2),
\label{gff}
\end{align}
which consists of displacement along the x-axis and a dynamic squeezing element that has been experimentally realized \cite{dynamic,sqff}. Then we have the corrected circuit to be 
\begin{align}
\widehat{T}_q = \widehat{F}_G \widehat{T}_1(q) = \hat{A}_q e^{ia_0\hat{x}^3}, 
\label{teff}
\end{align}
which is our required cubic gate along with a damping factor which is measurement dependent. This damping factor is very reminiscent of finite squeezing effects that appear in Gaussian continuous-variable cluster state quantum computation (see for example Eq. (177) of \cite{rmp}). Note that we have directly obtained the cubic gate with one instance of the GKP circuit and a suitable resource state.   \\

\begin{figure}
\begin{center}
\scalebox{0.75}{\includegraphics{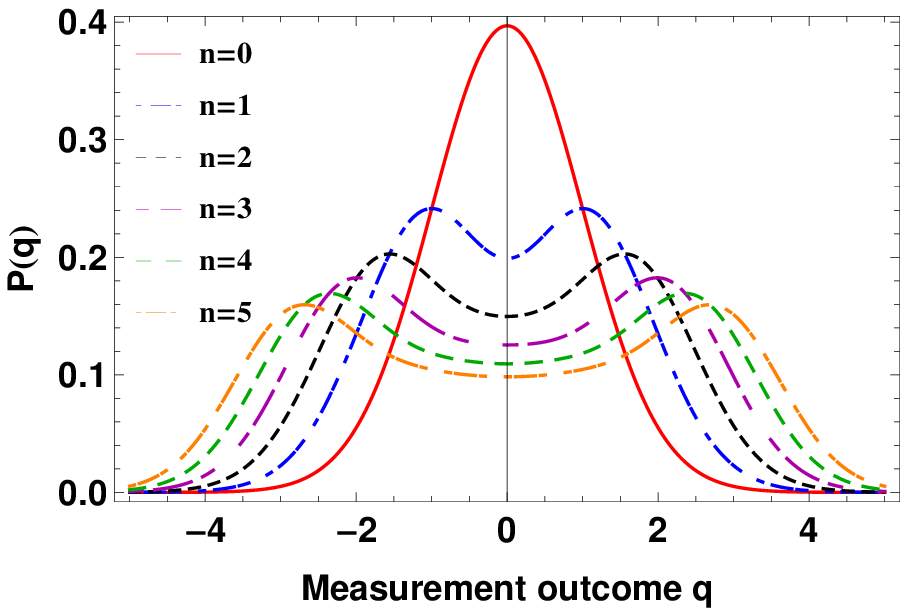}}
\end{center}
\caption{Probability of the homodyne detection $P(q)$ for input Fock states and target cubic gate strength set to $\gamma=0.1$, where the plots from the top to the bottom correspond to Fock states in increasing order  from $n=0$ to $n=5$. We observe that as the Fock number increases the probability distribution gets flatter.  }
\label{figb1}
\end{figure}

\begin{figure}
\begin{center}
\scalebox{0.75}{
\includegraphics{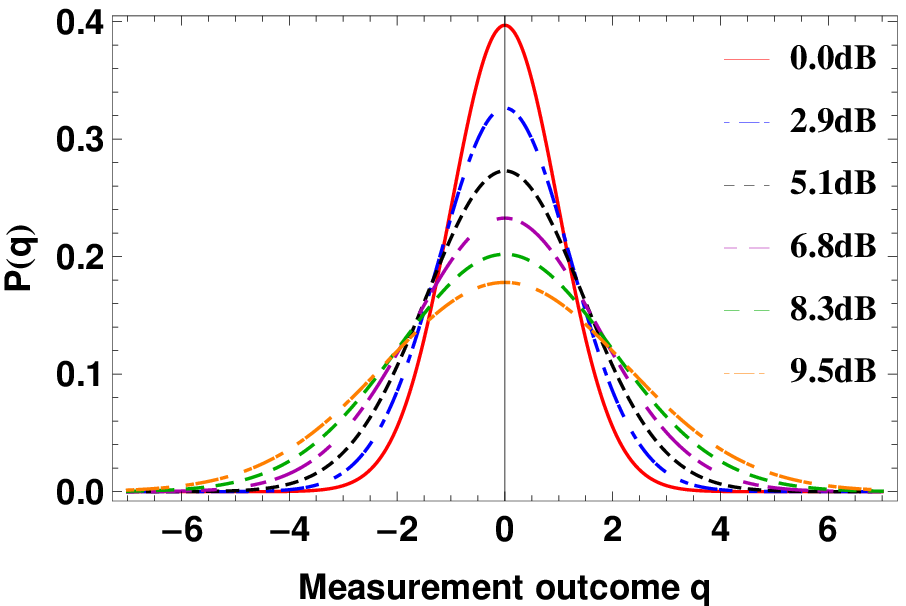}}
\end{center}
\caption{Probability of the homodyne detection $P(q)$ for input squeezed states and target cubic gate strength set to $\gamma=0.1$, where the plots from the top to the bottom correspond to squeezing value in increasing order  from $0$dB to $9.5$dB. We observe that higher input squeezing leads to a damping of the probability distribution.}
\label{figb2}
\end{figure}

\begin{figure}
\begin{center}
\scalebox{0.75}{
\includegraphics{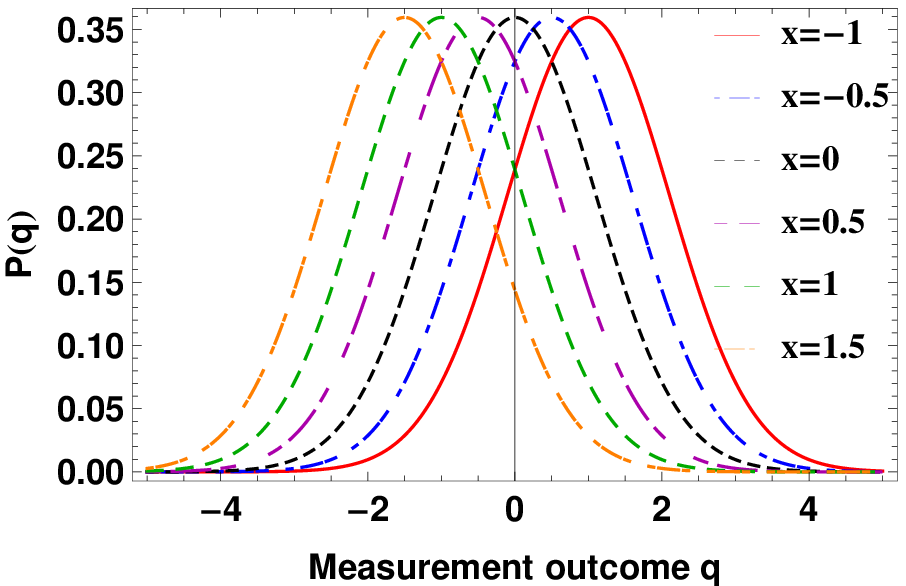}}
\end{center}
\caption{Probability of the homodyne detection $P(q)$ for input x-displaced vacuum states and target cubic gate strength set to $\gamma=0.1$, where the plots from right to left correspond to x-displacements ranging from $-1$ to $1.5$. We find that input displacements lead to proportional shifts for the probability distribution.}
\label{figb3}
\end{figure}

\noindent\textbf{Role of feed-forward corrections.} 
The effective operator $\widehat{T}_1(q)$ is probabilistically implemented and the probability depends on our fixed 03 resource state and the input test states as seen from Eq. \eqref{prob}. We plot this probability distribution for three classes of test states, namely, Fock states,  squeezed vacuum states, and coherent states respectively in Figs.\, \ref{figb1}, \ref{figb2}, \ref{figb3}. We find that for coherent states, the probability distribution is shifted by an amount equal to the x-displacement of the test state. For squeezed states the probability gets flattened with larger squeezing. A similar damping effect is observed for input test Fock states with regard to larger Fock numbers. 

These plots serve to better understand the role of feed-forward corrections and the probabilistic nature of the circuit. Let us briefly assume that we completely avoid the Gaussian feed-forward corrections, and just restrict the homodyne to a value $q_0 \pm \epsilon$, where $\epsilon$ accounts for some error we tolerate for the homodyne measurement. This would render the circuit non-deterministic as opposed to deterministic where all homodyne values are used with the need for suitable feed-forward corrections. In other words, we only look for events when the homodyne detection falls in our chosen range. For all homodyne detections within this value, we simply apply a pre-processing for the circuit by a displacement $X(q_0)$, where $X(t) = e^{-it\hat{p}}$. Note that $X(t) \hat{x}X(-t) = \hat{x}-t $. So we have the approximate effective circuit implementation to be 
\begin{align}
\widehat{T}_1(q_0) X(-q_0) \ket{\psi} &=  X(-q_0) \left(X(q_0) \widehat{T}_1(q_0) X(-q_0) \right) \ket{\psi}\nonumber\\ 
&= X(-q_0) \widehat{T}_1(0) \ket{\psi}.%, \nonumber\\ 
%\hat{T}_2 &= \hat{A}_0 \widehat{Z}(-1.5 a_0) e^{ia_0 \hat{x}^3}. 
\end{align}
If we post-process the circuit by two displacements $Z(3a_0/2) X(q_0) $ , we have the final effective operator to be given by 
\begin{align}
\widehat{T}_{0} &= Z(3a_0/2) X(q_0) X(-q_0)  \widehat{T}_1(0) \nonumber\\ &= \hat{A}_0 e^{ia_0 \hat{x}^3},
\end{align}
which is our target gate apart from the overall Gaussian `damping' factor. So at the cost of fixing the value of the homodyne detection, and thereby reducing the probability of success, we have gained that we do not need the feed-forward corrections, but only fixed pre and post circuit displacements. 

Suppose we fix the homodyne value to $q_0=0$ and $\epsilon = 10^{-2}$, for the test states in Figs. \ref{figb1} to \ref{figb3} we roughly incur a drop in probability of order $10^{-3}$. The conclusion is that depending on the test states and the quality of the optical elements in the feed-forward, one could in principle have the option of whether or not to fix the homodyne measurement. We however assume the more general case of requiring feed-forward when we present the optical circuit using 03 states. \\

\noindent
\textbf{Preparation of the 03 state.} For the preparation of the state we essentially follow the method presented in \cite{uptothree}. The first step is the preparation of a two-mode squeezed state (TMSS). This can be done with two single-mode squeezers pre- and post-processed by two beam splitters. Consider the single-mode squeezing operation that implements on the mode operators $R = (\hat{x}_1,\hat{p}_1)^T$ the transformation  
\begin{align}
S = {\rm diag}(e^{-r},e^{r}),
\end{align}
and similarly a squeezing $S^{-1}$ on a second mode \cite{note1}.
Let us sandwich this between two 50-50 beam splitters  that are inverses of each other where 
\begin{align}
BS(\pi/4) = \frac{1}{\sqrt{2}}\begin{pmatrix}
1\!\!1_2&1\!\!1_2\\
-1\!\!1_2&1\!\!1_2
\end{pmatrix},
\end{align}
where $1\!\!1_2$ is the $2\times 2$ identity matrix. 
Then we have that 
\begin{align}
BS(\pi/4)^{-1} (S \oplus S^{-1}) BS(\pi/4) = 
\begin{pmatrix}
c_r 1\!\!1_2& -s_r \sigma_3\\
-s_r \sigma_3 &c_r 1\!\!1_2\\
\end{pmatrix} \equiv S_2, 
\label{1to2}
\end{align}
where $c_r = \cosh{r}$ and $s_r = \sinh{r}$. This two-mode unitary squeeze operator $S_2$ acting on the vacuum state of two modes gives rise to  \begin{align}
\ket{\rm TMSS} = {\rm sech} r \, \sum_{n=0}^{\infty} (\tanh r)^n \ket{n} \ket{n}.
\end{align} But since the beam splitter leaves the vacuum state unchanged, we only need one beam splitter operation in Eq. \eqref{1to2}. 

The next step is to perform a three photon-subtraction  of the form
\begin{align}
\widehat{Y} = (\hat{a} + \beta_1) (\hat{a} + \beta_2) (\hat{a} + \beta_3) 
\end{align}
and then a measurement $\Pi_0 = \ket{0}\bra{0}$ on one arm of the TMSS. We drop the overall constant since $\widehat{Y}$ is a filter, label $\tanh r = y$, and we have 
\begin{align}
&\bra{0} 1\!\!1 \otimes \widehat{Y} \ket{\rm TMSS} \nonumber\\
&=  \beta_1 \beta_2 \beta_3 \ket{0} + y (\beta_1 \beta_2 + \beta_2 \beta_3 + \beta_3 \beta_1) \ket{1} + \nonumber\\ 
& ~~~~+ y^2 \sqrt{2} (\beta_1 + \beta_2 + \beta_3) \ket{2} + \sqrt{6} y^3 \ket{3}. 
\label{e32}
\end{align}
Setting $\beta_1 = c e^{i\pi/6}, \beta_2 = c e^{i 5\pi/6}, \beta_3 = c e^{i3\pi/2}$, we get the vector 
\begin{align}
i c^3 \ket{0} + \sqrt{6} y^3 \ket{3} \simeq \ket{0} + i (y/c)^3 \ket{3},
\end{align}
where we have absorbed the negative sign and the factor $\sqrt{6}$ into $c$. We have tunable parameters $c$ and $y(r)$ to obtain the desired 03 state. To implement the filter $\widehat{Y}$ we rewrite it as 
\begin{align}
\widehat{Y} = D(\beta_1)^{\dagger} \hat{a} D(\beta_1) D(\beta_2)^{\dagger} \hat{a} D(\beta_2) D(\beta_3)^{\dagger} \hat{a} D(\beta_3),  
\label{3psub}
\end{align}
where we have used the fact that $D(\beta)^{\dagger} \hat{a} D(\beta) = \hat{a} + \beta$. We can simplify the expression using $D(\beta) D(\alpha) = D(\alpha + \beta) e^{(\alpha \beta^* - \alpha^* \beta)/2}$ to obtain
\begin{align}
\widehat{Y} = \delta D(-\beta_1) \hat{a} D(\beta_1-\beta_2) \hat{a} D(\beta_2-\beta_3) \hat{a} D(\beta_3), 
\label{e35}
\end{align}
with all the scalars combined into $\delta$, which can be accounted for in the final normalization of the output state. We now consider implementation of photon subtraction and displacements needed for the operation in Eq. \eqref{e35}. \\

\noindent \textit{Photon subtraction element.} Consider a beam splitter $BS(\theta) = e^{\theta (\hat{a}^{\dagger} \hat{b} - \hat{a} \hat{b}^{\dagger})}$ with high transmissivity, i.e., $\theta <<1$. Then we can write $BS(\theta) = 1 + \theta (\hat{a}^{\dagger} \hat{b} - \hat{a} \hat{b}^{\dagger}) + O(\theta^2)$ so that 
\begin{align}
BS(\theta) \ket{\psi} \ket{0} &= \ket{\psi} \ket{0} - \theta \hat{a} \ket{\psi} \ket{1} + O(\theta^2).
\end{align}
If we measure a single photon count in the ancilla mode we obtain a photon subtraction on $\ket{\psi}$ as depicted in the circuit \begin{align} \label{psub}
&\mbox{
\Qcircuit @C=0.7em @R=2.5em { 
\lstick{\ket{\psi}}& \qw&\multigate{1}{BS} & \qw &\push{~\hat{a} \ket{\psi}} \qw &\\ 
\lstick{\ket{0}}& \qw&\ghost{BS} & \qw &\measureD{\mbox{$\Pi_n$}} & \cw ~~~~1 ~.
}}
\end{align}

%\begin{figure}
%\begin{center}
%\includegraphics{psub.eps}\\
%\includegraphics{disp.eps}
%\end{center}
%\caption{basic elements}
%\label{fig3}
%\end{figure}

\noindent \textit{Displacement element} \cite{dariano95,paris96}.
To implement a displacement operator we need a beam splitter with high transmissitivity and a large coherent state $\ket{z}$ in the environment as given by the optical circuit 
\begin{align}
\mbox{
\Qcircuit @C=1.2em @R=2.5em { 
\lstick{\ket{\psi}}&\qw&\multigate{1}{BS} &\qw &~~~D(\alpha(z))\ket{\psi}, &\\ 
\lstick{\ket{z}}&\qw&\ghost{BS} \qw &\qw &\measureD{\mbox{Tr}} 
}}
\end{align}
where ${\rm Tr}$ denotes the partial trace of the environment mode. We use the identity that $D(\alpha) \hat{\rho} D(\alpha)^{\dagger} = \lim_{|z| \to \infty} {\rm Tr}_2 (BS(\theta) \rho \otimes \ket{z} \bra{z} BS(\theta)^{\dagger})$, with $\alpha = i z \sin{\theta}$, $\theta << 1$, such that $z \sin{\theta} = {\rm constant}$.  \\

\begin{figure*}[ht]
\begin{center}
\scalebox{0.77}{
\includegraphics{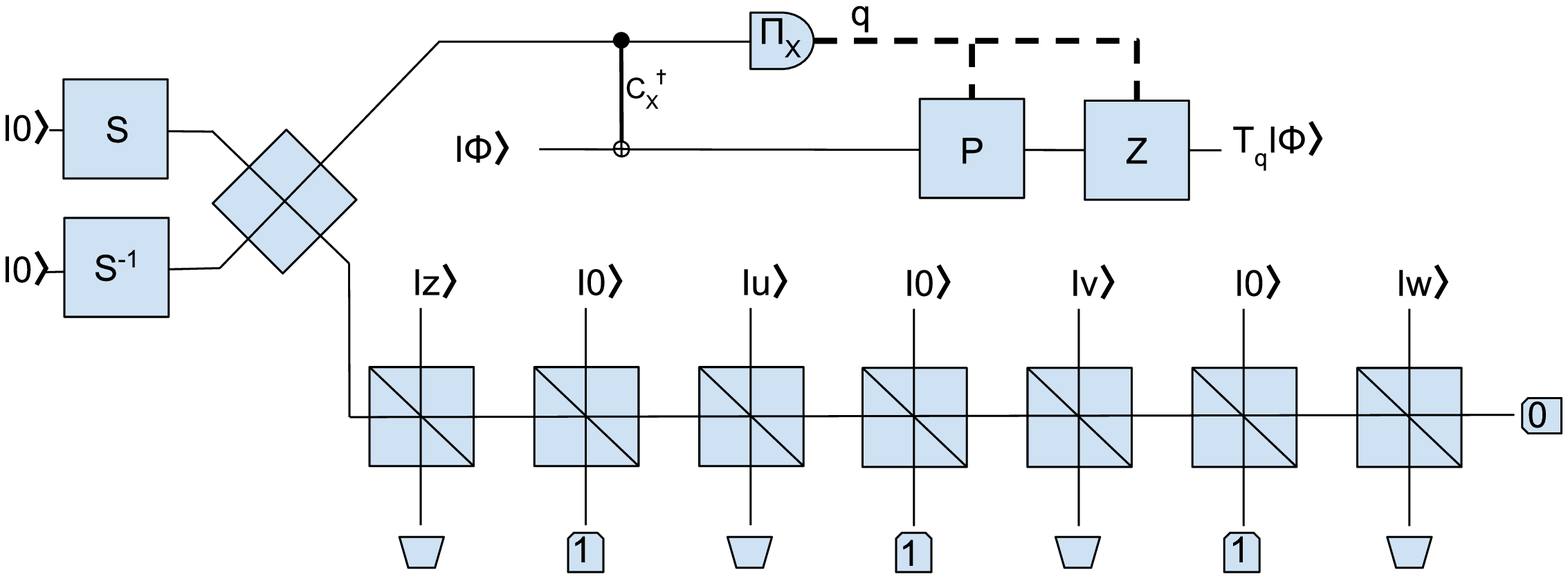}}
\end{center}
\caption{GKP circuit with the 03 resource state given by $c_a (\ket{0} + a \ket{3})$. The circuit implements a weak cubic phase gate up to an overall Gaussian noise factor $\widehat{T}_q = \hat{A}_q V(\gamma)$. The feed-forward (dashed lines) Gaussian gates are the quadratic phase gate $P$ and the displacement $Z$ both of which are conditioned on the homodyne measurement outcome $\Pi_x=q$, and are  therefore dynamic optical elements. The four coherent states $\ket{z}, \ket{u}, \ket{v}, \ket{w}$ are part of the displacement beams performed on one arm of the two-mode squeezed state as given in Eq. \eqref{e35}. The detector with label `$1$' corresponds to the measurement  $\{ \Pi_1, 1\!\!1 - \Pi_1\}$ and with `$0$'  to $\{ \Pi_0, 1\!\!1 - \Pi_0\}$. The gate $S$ stands for a single-mode squeeze operator. The entangling gate in the primary circuit is the $C_X^{\dagger}$ gate. 
}
\label{figd}
\end{figure*}

\noindent \textit{Photon-addition element.}  We first note the identity
\begin{align}
\label{arm}
(1\!\!1 \otimes \hat{a})\ket{TMSS} = (y \hat{a}^{\dag} \otimes 1\!\!1) \ket{TMSS}.
\end{align}
So instead of keeping the other arm of the TMSS on a long delay loop while the photon-subtraction operator of Eq. \,\eqref{3psub} is implemented, one could move some of the photon-subtraction elements to this arm using the identity in Eq. \eqref{arm}. The standard implementation of photon addition on an arbitrary state uses a beam splitter with high transmissivity and a single photon resource state as depicted in the optical circuit 
\begin{align}
    \mbox{
\Qcircuit @C=0.7em @R=2.5em { 
\lstick{\ket{\psi}}& \qw&\multigate{1}{BS} & \qw &\push{~\hat{a}^{\dag} \ket{\psi}} \qw &\\ 
\lstick{\ket{1}}& \qw&\ghost{BS} & \qw &\measureD{\mbox{$\Pi_n$}} & \cw ~~~0 ,
}}
\end{align}
to obtain the state before measurement as
\begin{align}
BS(\theta) \ket{\psi} \ket{1} &= \left[ 1\!\!1 +  \theta (\hat{a}^{\dagger} \hat{b} - \hat{a} \hat{b}^{\dagger} ) + O(\theta^2)\right]  \ket{\psi} \ket{1} \nonumber\\
&= \ket{\psi} \ket{1} + \theta \hat{a}^{\dagger}\ket{\psi} \ket{0} -\theta \hat{a}\ket{\psi} \ket{2} + O(\theta^2).
\end{align}
Then measuring the vacuum photon number on the ancilla mode implements a photon addition on $\ket{\psi}$.

However, in lieu of the single photon state one can instead implement photon addition using a weak two-mode squeeze operator $S_2(r) = e^{r(\hat{a}^{\dagger} \hat{b}^{\dagger} - \hat{a}\hat{b})}$, $r<<1$ \cite{hybrid}. We have 
\begin{align}
S_2(r) \ket{\psi} \ket{0} &= [1\!\!1 + r(\hat{a}^{\dagger} \hat{b}^{\dagger} - \hat{a}\hat{b})]\ket{\psi} \ket{0} \nonumber\\ 
&=  \ket{\psi} \ket{0} + r \hat{a}^{\dagger}\ket{\psi} \ket{1} + O(r^2).  
\end{align}
Detecting a single photon count in the second mode implements a photon addition onto the first mode as given in the circuit
\begin{align} \label{psub3}
&\mbox{
\Qcircuit @C=0.7em @R=2.5em { 
\lstick{\ket{\psi}}& \qw&\multigate{1}{S_2} & \qw &\push{~\hat{a}^{\dag} \ket{\psi}} \qw &\\ 
\lstick{\ket{0}}& \qw&\ghost{S_2} & \qw &\measureD{\mbox{$\Pi_n$}} & \cw ~~~~1~,
}}
\end{align}\\
analogous photon subtraction element in Eq. \eqref{psub}.  Finally using identity in Eq. \eqref{1to2} that relates a two-mode squeeze operator with single-mode squeeze operators and beam splitters,  the optical circuit in Eq. \eqref{psub3} becomes 
\begin{align}
&\mbox{
\Qcircuit @C=0.4em @R=2.5em { 
\lstick{\ket{\psi}}& \qw&\multigate{1}{BS} & \qw&\gate{S^{~}} &\multigate{1}{BS^{-1}}&\push{~\hat{a}^{\dag} \ket{\psi}} \qw &\\ 
\lstick{\ket{0}}& \qw&\ghost{BS} & \qw &\gate{S^{-1}}&\ghost{BS^{-1}}&\measureD{\mbox{$\Pi_n$}} & \cw ~~~1.
}} \nonumber
\end{align}

\noindent \textbf{Gate and detector count.} We finally put all the optical elements in place as depicted in Fig. \ref{figd} so that we are able to make a detailed gate and detector count. We note that the QND entangling operation in the GKP circuit requires two single-mode squeezers and two beam splitters \cite{qnd,cvlinear} sandwiched between Fourier operators which are specific phase shifters. The quadratic feed-forward requires two phase shifters and a squeezing element \cite{dynamic,sqff}, along with a displacement element as shown in Eq. \eqref{gff}. The state preparation requires eight beam splitters, two single-mode squeezers, three single photon resolving detectors, and one vacuum projection. The homodyne detection of the GKP circuit requires one beam splitter. To sum up,  all the non-Gaussianity from the resource state that is transferred into the circuit comes about exactly through three single photon detectors. \\

\begin{table}
\begin{tabular}{l|c}
\hline
\textbf{Optical element} &{\bf Count} \\
\hline
\hline
Beam splitters (BS)&12 \\
Squeezers (S) & 5\\ 
Single photon resolving detectors & 3 \\
APD &1\\
Phase shifters &4 \\
Dynamic gates & 2 \\
\hline
\end{tabular}
\caption{Gate and detector count of the total optical circuit when using the 03 state as a resource in the GKP circuit. APD stands for avalanche photo diode and is also known as the on/off detector or the single photon counter and corresponds to the measurement $\{\Pi_0 =\proj{0}, 1\!\!1 - \Pi_0 \}$. The single photon resolving detectors correspond to the measurement $\{\Pi_1 =\proj{1}, 1\!\!1 - \Pi_1 \}$. We call gates that are conditioned on the homodyne measurement outcome in the circuit as dynamic gates. }
\label{table1}
\end{table}

%\begin{figure}[ht]
%\begin{center}
%\includegraphics{padd12.eps}
%\end{center}
%\caption{Using a two-mode squeezer for photon-addition}
%\label{fig2}
%\end{figure}

\noindent \textbf{Squeezed effective operator.} 
If the input test states contain a squeezing operation, this in turn can be interpreted as replacing the original effective operation by a squeezed operation, i.e., 
\begin{align}
\widehat{T} \left(  S(r) \ket{\psi} \right) &= S(r) \left( S(r)^{\dagger} \widehat{T} S(r) \right) \ket{\psi} \nonumber\\
&= S(r) \widehat{T}(r) \ket{\psi}.
\end{align}
In the case of the 03 resource state we have that 
\begin{align}
\widehat{T}_q(r) = \exp[-\frac{1}{2}\left(\hat{x}/r +q\right)^2 + i a_0 (\hat{x}/r)^3 ], 
\end{align}
where $S(r)^{\dagger} \hat{x} S(r) = \hat{x}/r$. Let us assume $r >1$.  
The effects of squeezing are opposing: on the one hand this may inhibit the role of the Gaussian damping factor while at the same time it also reduces the strength of the cubic phase gate. So having \textit{a priori} information about the test states one could tune the effective operator that is to be implemented to improve the output fidelities. Also, another  way to use squeezing to mitigate the effect of the damping operator would be to squeeze the resource state while simultaneously tuning the coefficient $a_0$ in the 03 state to produce the same gate action.  \\

\noindent \textbf{Cubic phase gate to higher-order in accuracy.}
Let us consider a general unitary operator $U_H = e^{iHt} = \cos[Ht] + i \sin [Ht]$. For small enough evolution times we have the expansion to second order in time to be given by 
\begin{align}
U = 1\!\!1  +i Ht -\frac{H^2t^2}{2} +O(t^3). 
\end{align}
Due to the finite-order expansion we lose unitarity. Such expansions would appear when we concatenate cubic gates to generate higher-order gates. To achieve the required accuracy using the resource state method we need to prepare a very intricate state that has a superposition of the vacuum and up to the six photon excited state, that is at present experimentally out of reach. 

As an approximation to this second order expansion we consider the product expansion of the unitary operator. It is well known that the exponent of a matrix $X$ can be written as a limit of products given by 
\begin{align}
e^X = \lim_{n \to \infty} \left( 1 + \frac{X}{n}\right)^n. 
\end{align}
If we consider this product for $n=2$ and set $X=iHt$,  we obtain 
\begin{align}
\left(1+\frac{iHt}{2}\right)^2 = 1\!\!1 + iHt  - \frac{H^2 t^2}{4}.  
\end{align}
The difference between the Taylor expansion and the product expansion to the second order is given by 
\begin{align}
\triangle_2 = \frac{H^2t^2}
{4}.
\end{align}
With regard to the cubic gate we have that $H = \hat{x}^3$ and so $\triangle_2 =\hat{x}^6 t^2/4$. 

 \begin{figure*}
\begin{center}
\scalebox{0.665}{\includegraphics{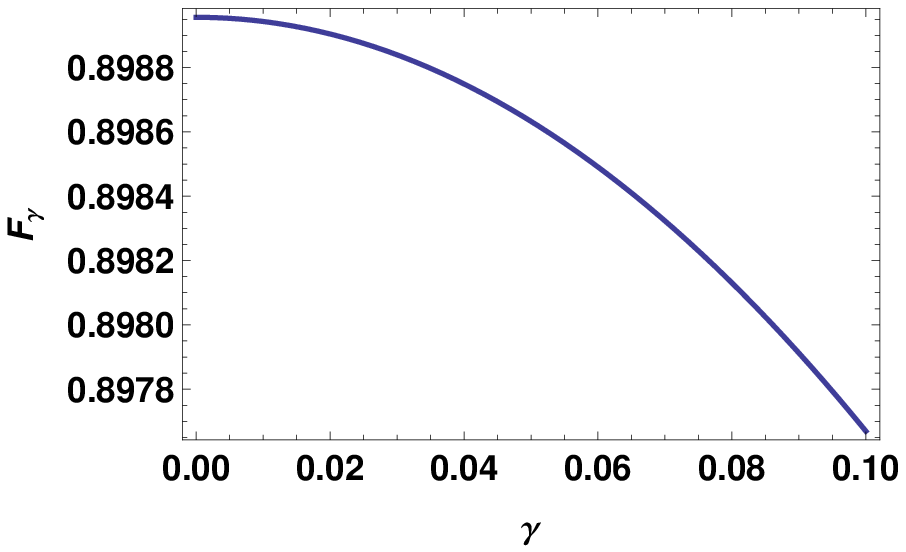}}
\scalebox{0.615}{\includegraphics{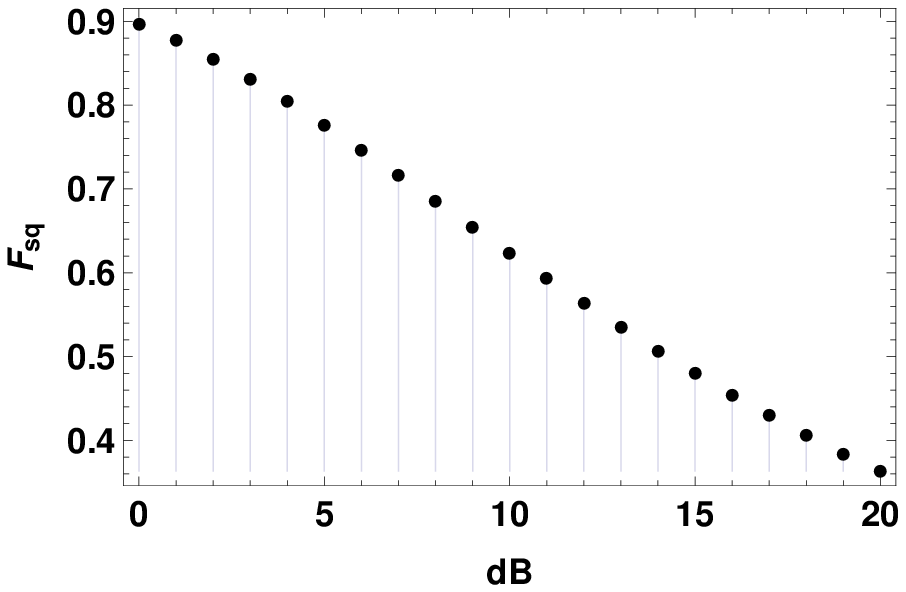}}
\scalebox{0.615}{\includegraphics{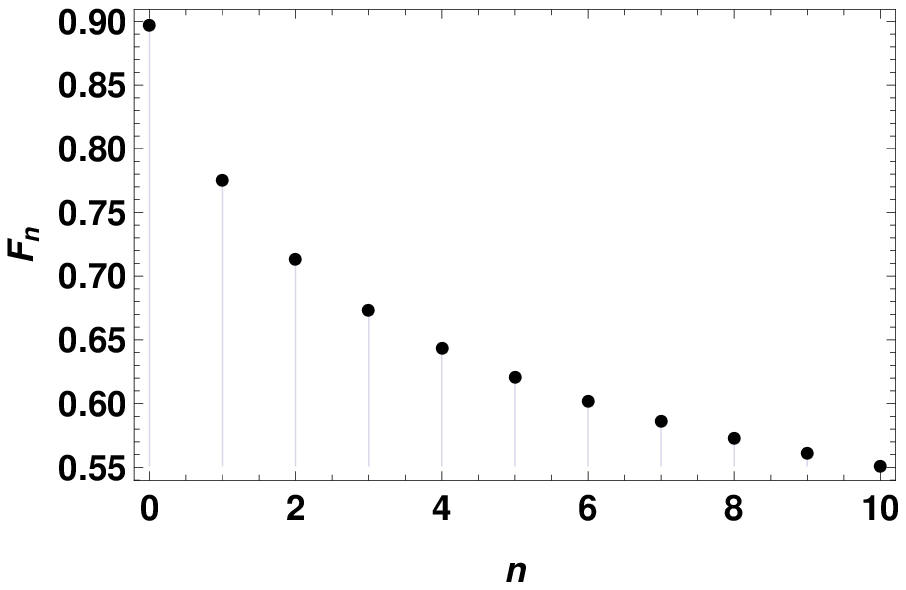}}
\end{center}
\caption{Gate fidelity $F_{\gamma}, F_{sq}, F_{n}$ using respectively (a) coherent states, as a function of target cubic gate strength $\gamma$ (since the fidelity is displacement independent for a fixed $\gamma$), (b) squeezed states, for $\gamma =0.1$, and (c) Fock states, for $\gamma=0.1$.}
\label{figb}
\end{figure*}

Finally, to make use of the product form for approximating the cubic gate to second order, we would need to implement the GKP circuit with the 03 state twice. In this way one can in principle go to higher-order accuracy of the cubic phase gate using up more copies of the 03 state. This could possibly be a better strategy to approximate higher-order cubic gates, with due care given to the approximations, as compared to creating very complex resource states that require creating and maintaining superpositions with multiple Fock states.  \\

\subsection{Gate performance and the effect of the Gaussian noise operator $\hat{A}_q$}
 There are two commonly used notions of fidelity for a quantum process of finite-dimensional systems, the worst case and average case fidelity \cite{nielsen}. While the worst case fidelity can be easily generalized and indeed such measures have been used for benchmarking \cite{sqbench}, the average case cannot be directly generalized due to the non-existence of a Haar measure for continuous-variable systems \cite{blume}. We take motivation from Ref. \cite{pariscloning} to define suitable gate fidelities that are similar in spirit to the average fidelities.
 
 The fidelity between two states is defined as $F(\ket{\psi},\ket{\phi}) = |\braket{\psi}{\phi}|$. So we have that the gate fidelity of any operator $\widehat{T}$ that approximates the cubic phase gate $V(\gamma)$ on a test state $\ket{\psi}$ to be given by 
\begin{align}
 F \left(\frac{\widehat{T}\ket{\psi}}{||\widehat{T}\ket{\psi}||},V(\gamma)\ket{\psi}\right),
\end{align}
where $||\ket{v}|| = \sqrt{\braket{v}{v}}$. The normalization is needed because in general the approximate operator is a filter, i.e., not trace-preserving.

In our case of using the 03 resource state and in the weak cubic regime, i.e. $\gamma <<1$, we have the effective operator post feed-forward correction to be the one given in Eq. \eqref{teff}. Here the effective operator $\widehat{T}_q=\hat{A}_q V(\gamma)$ differs from the cubic gate only by an overall Gaussian noise factor which is dependent on the outcome of the homodyne measurement. So we have the fidelity on a given test state to be 
\begin{align}
\label{fpsi}
F_{q,\psi}  &= F \left(\frac{\widehat{T}_q \ket{\psi}}{||\widehat{T}_q \ket{\psi}||},V(\gamma)\ket{\psi}\right) \nonumber\\
&= \frac{\bra{\psi} \hat{A}_q \ket{\psi}}{\bra{\psi} \hat{A}_q^2 \ket{\psi}^{1/2}} =  \frac{\int dx |\psi(x)|^2 A_q(x)}{[\int dx |\psi(x)|^2 A_q(x)^2]^{1/2}},  
\end{align}
where we used the action of $\hat{A}_q$ from Eq. \eqref{aqx}. Since the homodyne outcome is probabilistic with weight given in Eq. \eqref{prob} we can average over these outcomes to obtain 
\begin{align}
F_{\psi,\gamma} = \int dq \, p(q,\gamma) F_{q,\psi}, 
\label{psiprob}
\end{align}
where $\gamma$ is the parameter of the target cubic gate. 

We can equivalently define the figures of merit in terms of trace distance in lieu of  fidelity and with suitable modifications. The relation connecting trace distance and fidelity between pure states is 
$D(|\psi\rangle,|\phi\rangle) =
\sqrt{1-|\bra{\phi}{\psi}\rangle|^2} = \sqrt{1-F(\ket{\psi},\ket{\phi})^2}$ \cite{chuang}. We now consider the role of damping on important test states such as coherent states, squeezed vacuum states, and Fock states. \\

\noindent \textbf{Examples.} \textit{Coherent states.} As our first example we consider coherent states $\ket{\alpha}$. A moment's notice at Eq. \eqref{fpsi} informs one that displacements along the momentum axis does not enter the expressions for fidelity. So we only consider coherent states that are displaced along the $x$ axis and we write its wave-function as 
\begin{align}
\langle x\ket{\alpha(x_0)} = \frac{1}{\pi^{1/4}} e^{-(x-x_0)^2/2}. 
\end{align}
In this case one finds that 
\begin{align}
F_{q,x_0} = \left[\frac{2\sqrt2}{3}\right]^{1/2}  e^{-(q+x_0)^2/12}.
\end{align}
One can then compute the average in Eq. \eqref{psiprob} to obtain $F_{x_0,\gamma}$. It turns out that $F_{x_0,\gamma}$ is independent of $x_0$, i.e., all coherent states give the same fidelity, as can be verified from the integrals. So we plot $F_{\gamma} = F(x_0,\gamma)$ as a function of $\gamma$ in the first plot of Fig. \ref{figb}. We find that the role of damping is only marginal and that the fidelity is close to $0.9$ for the entire range of $\gamma  \in [0,0.1]$. Note that there is a damping factor even for $\gamma =0$ due to the choice of the resource state. \\
%\begin{figure}[ht]
%\begin{center}
%\includegraphics[width=\columnwidth]{fid-coh.eps}
%\end{center}
%\caption{coherent state fidelity as a function of cubic strength. All coherent states have the same fidelity. The fidelity captures the role of the damping factor in the low cubic phase gate strength. }
%\label{fid-coh}
%\end{figure}

\begin{table*}
\begin{tabular}{|l|cccc|c|c|c|}
\hline
\backslashbox{Property}{Scheme} &GKP&MFF&ON+GKP&ATF1&ATF2& MPSW&AnGM\\
\hline
\hline
Resource state (RS)&$~\bra{n} Z(w)S_2\ket{00}$&$S(1+i\chi \hat{x}^3)\ket{0}$&$\ket{0} +ia\ket{3}$&$\hat{a} \ket{\alpha,r}$&$\ket{\alpha,r}$&$\ket{\alpha}$&$\ket{0}+\beta \ket{1} + \gamma \ket{2} + \delta\ket{3}$\\
Fixed resource &\cmark&\cmark&\cmark&\xmark&\xmark&\cmark&\cmark\\
Type of resource &nG&nG&nG&nG&G&G&nG\\
Entangling gate&$C_X^{\dag}$&$C_X^{\dag}$&$C_X^{\dag}$&$C_Z$&$C_Z$&$C_{\beta}$&beam splitter\\
Measurement (M)&$\Pi_x$&$\Pi_x$&$\Pi_x$&$\Pi_p$&$\Pi_n$&$\Pi_{0/1}$&adaptive heterodyne\\
Source of nG &RS&RS&RS&RS&M&M&RS\\
Implementation&approximate&1\ts{st} order&1\ts{st} order&monomial&monomial&monomial&1\ts{st} order\\
Feed-forward&P, X/Z&P, X/Z&P, X/Z&P, X/Z&X/Z&None&X/Z\\
Deterministic&\cmark&\cmark&\cmark&\cmark&\xmark&\xmark&\cmark\\
Circuit repetition&1&1&1&3&3&3&1\\
\hline
\end{tabular}
\caption{Showing a comparison of the seven schemes for implementing the cubic phase gate. $S$ denotes a single-mode squeeze operator, $S_2$ the two-mode squeeze operator, and $\beta,\gamma,\delta$ are fixed constants. G/nG stands for Gaussian/non-Gaussian(ity). With regard to measurements $\Pi_x$ corresponds to a homodyne measurement of quadrature $\hat{x}$, $\Pi_p$ to homodyne measurement of $\hat{p}$, $\Pi_n$ to a photon number resolving detection, $\Pi_{0/1}$ to the measurement $\{\Pi_0= \proj{0}, \Pi_1 =\proj{1}, 1\!\!1 - \Pi_0 - \Pi_1 \}$. Gates $P,X,Z$ denote the quadratic phase gate, x-displacement and p-displacement, respectively. The row `Source of nG' denotes whether the non-Gaussianity in the circuit originated from the resource state (RS) or the measurement (M). Entangling gates $C_X, \, C_Z,\,C_{\alpha}$ denote the controlled-X, controlled-Z, and the controlled-$\alpha$ gate. The implementation refers to the final form of the effective operation  as to whether it is an approximate cubic gate, the cubic gate to first order in accuracy (or a weak cubic gate), or if it implements a monomial operator in $\hat{x}$. The row `Deterministic' denotes whether the computation uses all measurement results or only certain results through post-selection, i.e. whether the method is deterministic or not. }
\label{table2}
\end{table*}

\noindent\textit{Squeezed vacuum and Fock states.} We repeat the calculation of fidelity for squeezed states and Fock states as test states. We plot the effect of damping respectively in plots 2 and 3 of Fig. \ref{figb}. We find that the fidelity decreases with increase in either squeezing or the Fock state number. If we had an apriori distribution for these test states, we can readily obtain the corresponding average fidelity.\hfill $\square$
%\begin{figure}[ht]
%\begin{center}
%\includegraphics[width=\columnwidth]{fid-fock.eps}
%\end{center}
%\caption{For a fixed $\gamma=0.1$, fidelity of fock states. we see it drops rapidly. }
%\label{fid-fock}
%\end{figure}
%\begin{figure}[ht]
%\begin{center}
%\includegraphics[width=\columnwidth]{fid-sq.eps}
%\end{center}
%\caption{For a fixed $\gamma=0.1$, fidelity of squeezed states. we see it first increases slightly and then starts falling off. }
%\label{fid-sq}
%\end{figure}

%\noindent Finally, one could define a worst case fidelity as 
%\begin{align}
%F_{\rm min}(T_{\rm eff}) &= \min_{\psi} F \left(\frac{T_{\rm eff} \ket{\psi}}{||T_{\rm eff} \ket{\psi}||},\Gamma_{\nu}\ket{\psi}\right).
%\end{align}
%where $F(\ket{\psi_1},\ket{\psi_2}) = |\braket{\psi_1}{\psi_2}|$, but we pursue this elsewhere. 

\subsection{Quartic gate}
 Even though the cubic gate is sufficient for universal quantum computation, there are two primary reasons for which one needs to consider higher-order nonlinear gates. The first is that it requires six cubic gates to implement an approximate quartic gate and this number grows rapidly for higher-order gates. The second reason is that one requires a cubic gate to at least second order in accuracy to implement a quartic gate to first order \cite{mff,horder}. These two considerations give a possible trade-off between constructing higher accuracy cubic gates or directly obtaining higher-order gates. We now briefly describe a proof-of-principle construction of a quartic gate from our general ON state method which we now take to be the 04 state. \\

\noindent \textbf{GKP circuit with the 04 state.} 
We now consider the 04 state as a resource state and define it as $\ket{\chi_a} = d_a (\ket{0} + a \ket{4})$, where $d_a = (1+|a|^2)^{-1/2}$. To use it as a resource state in the GKP circuit we write down the position wave function  which is 
\begin{align}
\chi_a(x) &= d_a \pi^{-1/4} e^{-x^2/2} \left[H_0(x) + \frac{a}{\sqrt{2^4 4!}} H_4(x) \right] \nonumber\\ 
&= d_a \pi^{-1/4} e^{-x^2/2} \left[1 + \frac{16a}{\sqrt{2^4 4!}} (x^4 - 3x^2 + 3/4) \right]
\end{align}
Let us set $a = i \sqrt{3/2} a_0 $ and we have 
\begin{align}
\chi_a(x) &= d_a \pi^{-1/4} e^{-x^2/2} [1 + i a_0 (x^4 - 3x^2 +3/4) ]. 
\end{align}
Then if we input this state into the GKP circuit we have the effective filter operator post-homodyne measurement outcome $q$ to be given by 
\begin{align}
\widehat{T}_2(q) = e^{-(\hat{x}+q)^2/2} [1\!\!1 + i a_0 ((\hat{x}+q)^4 - 3(\hat{x}+q)^2 +3/4)].
\end{align}
The above operator can in turn be interpreted as a first-order expansion for a unitary operator which would hold true if $a_0<<1$. So we have that 
\begin{align}
\widehat{T}_2(q) &= e^{-(\hat{x}+q)^2/2} e^{ia_0 ((\hat{x}+q)^4 - 3(\hat{x}+q)^2 +3/4)}.
\end{align}
We drop the overall phase factor and combine the Gaussian terms to obtain 
\begin{align}
\widehat{T}_2(q) &= e^{-\beta (\hat{x}+q)^2/2} e^{ia_0 (\hat{x}+q)^4}, \beta = (1+ i 6a_0). 
\end{align}
It is now clear that if we want to apply the exact quartic gate we would need corrections which are themselves cubic in nature as can be seen by expanding the Hamiltonian of the corresponding unitary. Again using the explicit form of the probability distribution of the homodyne measurement one may be able to simplify the effective circuit. We however explore the quartic gate in more detail elsewhere. \\

%\section{Targeting gates using approximate resources}

\section{Discussion}
In this work we introduced a new class of states called ON states and outlined how they could be used as a unit resource imperative in the GKP circuit to implement nonlinear gates to first order. The first order implementation approximates the unitary phase gate when the gate strength is small. A road map to future works include increasing the accuracy of the cubic gate, and  implementing higher-order gates. This article therefore gives a strong motivation for focusing on improved ON state preparation. 

We performed a complete analysis of the optical implementation starting from the resource state preparation, homodyne measurement analysis, the role of feed-forward, and the final gate fidelities that capture the role of the unavoidable Gaussian noise in the circuit. We find that the homodyne measurement probability distribution gets damped with higher number in the Fock state or higher dB of squeezing, and the distribution gets translated with respect to input coherent states. The gate fidelities also drop with respect to increased squeezing and Fock number, but remains invariant under input displacements of the coherent states. 

Table. \ref{table1} detailed the number of units required of each type of optical element since this will provide an insight into the circuit depth and will also help to understand losses that is very crucial from an experimental point of view.  We also list a comparison of properties of various implementations of the cubic phase gate in Table. \ref{table2}. We hope that this investigation would eventually lead to optimizing the total optical circuit by picking out the best features in the various methods to improve gate fidelity and success probability, since there does not seem to be a best candidate. For example, from Table \ref{table2}, we see that the methods  ATF2, MPSW, and AnGM, do not require a quadratic feed-forward. 

The reason the cubic phase gate has attracted much attention is due to the fact that it is the lowest order non-linear quadrature phase gate. Its implementation is a major challenge one needs to overcome in the physical medium of photonics to truly exploit the full potential of universal quantum computation. We believe that our present work provides a fresh perspective on this problem as well as placing it in context with respect to previous works. This allows us to increase the quality and the success probability of higher-order nonlinear gates, and understand the scope of circuit flexibility.  With a steady on-demand supply of the unit resource states in conjunction with improved photon-subtraction methods \cite{purifyps,ps2}, the circuit can be scaled up with repeated use to improve both accuracy and order.

\noindent \textit{Acknowledgments:}
We wish to thank  Brajesh Gupt, Daiqin Su, and Zachary Vernon for useful discussions. 

%\textbf{Could also cite the Cerf paper on entropy production in non-Gaussian states and the PRX error correction paper which used ON states as the first code word in binomial codes}


\begin{thebibliography}{}
\bibitem{note1} In this section alone we interchangeably use $S$ to denote both the symplectic transformation and the single-mode squeeze operator, and similarly $BS$ for the beam splitter unitary and the symplectic element. In general $S$ and $BS$ stand for the single-mode squeeze gate and the beam splitter gate respectively.
\end{thebibliography}
\end{document}